\documentclass[aps,prl,twocolumn]{revtex4}
\usepackage{amssymb,graphicx}
\usepackage{times}
\newcommand{\half}{\frac{1}{2}}

\newcommand{\fig}[2]{\includegraphics[width=#1]{./figures/#2}}
\newcommand{\Fig}[1]{\includegraphics[width=\columnwidth]{./figures/#1}}
\newlength{\bilderlength}

\newcommand{\rme}{{\mathrm{e}}}
\newcommand{\rmd}{{\mathrm{d}}}

\begin{document}
\bibliographystyle{KAY}
\title{\sffamily \bfseries \Large Random field spin models beyond one
loop: a mechanism for decreasing the lower critical dimension}
\author{\sffamily\bfseries\normalsize Pierre Le Doussal and Kay J\"org
Wiese \vspace*{3mm}} \affiliation{ CNRS-Laboratoire de Physique
Th{\'e}orique de l'Ecole Normale Sup{\'e}rieure, 24 rue Lhomond, 75005
Paris, France. } \date{\small\today}
\begin{abstract}
The functional RG for the random field and random anisotropy $O(N)$
sigma-models is studied to two loop.  The ferromagnetic/disordered
(F/D) transition fixed point is found to next order in $d=4+\epsilon$
for $N > N_c$ ($N_c=2.8347408$ for random field, $N_c=9.44121$ for
random anisotropy). For $N < N_c$ the lower critical dimension
$d=d_{\mathrm{lc}}$ plunges below $d_{\mathrm{lc}}=4$: we find {\it
two} fixed points, one describing the quasi-ordered phase, the other
is novel and describes the F/D transition.  $d_{\mathrm{lc}}$ can be
obtained in an $( N_c-N)$-expansion. The theory is also analyzed at
large $N$ and a glassy regime is found. 
\end{abstract}
\maketitle

It is important for numerous experiments to understand how the
spontaneous ordering in a pure system is changed by quenched substrate
impurities. One class of systems are modeled by elastic objects in
random potentials (so-called random manifolds, RM). Another class are
$O(N)$ classical spin models with ferromagnetic (Ferro) couplings in
presence of random fields (RF) or anisotropies (RA). The latter
describe amorphous magnets \cite{HarrisPlischkeZuckermann1973}.
Examples of RF are liquid crystals in porous media
\cite{FeldmanPelcovits2004}, He-3 in aerogels \cite{Matsumoto1997},
nematic elastomers \cite{FridrikhTerentjev1997}, and ferroelectrics
\cite{Feldman2001}.  The XY random field case $N=2$ is common to both
classes and describes periodic RM such as charge density waves, Wigner
crystals and vortex lattices
\cite{BlatterFeigelmanGeshkenbeinLarkinVinokur1994}.  Larkin showed
\cite{Larkin1970} that the well-understood pure fixed points (FP) of
both classes are perturbatively unstable to weak disorder for
$d<d_{\mathrm{uc}}$ ($d_{\mathrm{uc}}=4$ in the generic case). For a
continuous symmetry (i.e.\ the RF Heisenberg model) it was proven
\cite{AizenmanWehr1989} that order is destroyed below $d=4$.  This
does not settle the difficult question of the lower critical dimension
$d_{\mathrm{lc}}$ as a weak-disorder phase can
 survive below $d_{\mathrm{uc}}$, if associated to a non-trivial
FP, as predicted in $d=3$ for the Bragg-glass phase with quasi
long-range order (QLRO) \cite{GiamarchiLeDoussal1995}.  For the random
field Ising model $N=1$ (RFIM) it was argued \cite{ImryMa1975}, then
proven \cite{Imbrie1984+BricmontKupiainen1987} that the ferromagnetic
phase survives in $d=3$.  Developing a field theory to predict
$d_{\mathrm{lc}}$,  and the exponents of the weak-disorder phase
and the Ferro/Disordered (F/D) transition, has been a long-standing
challenge. Both extensive numerics and experiments have not yet
produced an unambiguous picture.  Among the debated issues are the
critical region of the 3D RFIM
\cite{BelangerBookYoung+NattermannBookYoung} and the possibility of a
QLRO phase in amorphous magnets
\cite{BarbaraCouachDieny1987,Fisch1998,Feldman2001}.

A peculiar property shared by both classes is that observables are
identical to all orders to the corresponding ones in a $d-2$ thermal
model
\cite{AharonyImryShangkeng1976+EfetovLarkin1977+ParisiSourlas1979}.
This dimensional reduction (DR) naively predicts $d_{\mathrm{lc}}=4$
for the weak-disorder phase in a RF with a continuum symmetry
\cite{16} and no Ferro order for the $d=3$ RFIM, which is proven wrong
\cite{Imbrie1984+BricmontKupiainen1987}. It also predicts
$d^{\mathrm{t}}_{\mathrm{uc}}=6$ for the F/D {\it transition} FP.
While there is agreement that multiple local minima are responsible
for DR failure, constructing the field theory beyond DR is a
formidable challenge.  Recent attempts include a reexamination of the
$\phi^4$ theory (i.e.\ soft spins) for the F/D transition near $d=6$
\cite{BrezinDeDominicis}.  Previous large-$N$ approaches failed to
find a non-trivial FP, but a self-consistent resummation including the
$1/N$ corrections hinted at exponents different from DR (without
succeeding in computing them) from a solution breaking replica
symmetry \cite{MezardYoung1992}.

As for the pure $O(N)$ model, an alternative to the soft-spin version
(near $d=6$) is the sigma model near the lower critical dimension
(here presumed to be $d=4$).  In 1985 D.S. Fisher \cite{Fisher1985b}
noticed that an infinite set of operators become relevant near $d=4$
in the RF $O(N)$ model. These were encoded in a single function
$R(\phi)$ for which Functional RG equations (FRG) were derived to one
loop, but no new FP was found.  For a RM problem \cite{DSFisher1986}
it was found that a cusp develops in the function $R(\phi)$ (the
disorder correlator), a crucial feature which allows to obtain
non-trivial exponents and evade DR. A fixed point for the RF model was
later found \cite{GiamarchiLeDoussal1995} in $d=4-\epsilon$ for $N=2$.
It was noticed only very recently \cite{Feldman2002} that the 1-loop
FRG equations of Ref.\ \cite{Fisher1985b} possess fixed points in
$d=4+\epsilon$ for $N\ge 3$, providing a description of the
long-sought critical exponents of the F/D transition.

In spite of these advances, many questions remain. Constructing FRG
beyond one loop (and checking its internal consistency) is highly
non-trivial. Progress was made for RM
\cite{ChauveLeDoussalWiese2000a,LeDoussalWiese2001}, and one hopes
for extension to RF. Some questions necessitate a 2-loop
treatment, e.g.\ for the depinning transition, as shown in
\cite{LeDoussalWieseChauve2002}. In RF and RA models the 1-loop
analysis predicted some repulsive FP in $d=4+\epsilon$ (for larger
values of $N$), and some attractive ones
\cite{GiamarchiLeDoussal1995,Feldman2000} in $d=4-\epsilon$. The
overall picture thus suggests a lowering of the critical dimension,
but how it occurs remains unclear.  Finally the situation at large $N$
is also puzzling.  Recently, via a truncation of exact RG
\cite{TarjusTissier2004} it was claimed that DR is recovered for $N$
large.

Our aim in this Letter is twofold. We reexamine the overall scenario
for the fixed points and phases of the $O(N)$ model using FRG. This
requires the FRG to two loop. Here we present selected results,
details are presented elsewhere \cite{LeDoussalWieseToBePublished}. We
find a novel mechanism for how the lower critical dimension is
decreased below $d=4$ for $N<N_c$ at some critical value $N_c$.  We
obtain a description of the bifurcation which occurs at $N_c$, and
below $N_c$ we find {\em two} perturbative FPs. Thanks to 2-loop terms
$d_{\mathrm{lc}}$ can be computed in an expansion in $N_c-N$, and the
Ferro/Para FP {\em below} $d=4$ is found. A study at large $N$
indicates that some glassy behavior survives there.

Let us consider $O(N)$ classical spins $\vec{n}(x)$ of unit norm
$\vec{n}^2=1$. To describe disorder-averaged correlations one
introduces replicas $\vec{n}_a(x)$, $a=1,\dots, k$, the limit $k=0$
being implicit everywhere. The starting model is a non-linear sigma
model, of partition function ${\cal Z}=\int {\cal D} [\pi]\,
\rme^{-{\cal S}[\pi]}$ and action:
\begin{eqnarray}\label{action}
 {\cal S}[\pi ] = \int \rmd^d x &\Big[&\! \frac{1}{2 T_{0}} \sum_a [ (\nabla 
\vec \pi_a)^2 + (\nabla \sigma_a)^2 ] - \frac{1}{T_{0}} \sum_a M_0 \sigma_a 
\nonumber  \\
&& - \frac{1}{2 T_0^2} \sum_{a b} \hat R_0(\vec n_a \vec n_b) \Big]\ ,
\end{eqnarray}
where $\vec n_a=(\sigma_a,\vec \pi_a)$ with $\sigma_a(x)=\sqrt{1-\vec
\pi_a(x)^2}$. A small uniform external field $\sim M_0 (1,\vec 0)$
acts as an infrared cutoff. Fluctuations around its direction are
parameterized by $( N-1)$ $\pi$-modes. The ferromagnetic exchange
produces the 1-replica part, while the random field yields the
2-replica term $\hat R_0(z) = z$ for a bare Gaussian RF. RA
corresponds to $\hat R_0(z) = z^2$.  As shown in \cite{Fisher1985b}
one must include a full function $\hat R_0(z)$, as it is generated
under RG. It is marginal in $d=4$.

To obtain physics at large scales, one computes perturbatively the
effective action $\Gamma[n_a(x)]$. It can be expanded in gradients
near a uniform background configuration $n_a^0$, and splitted in 1-,
2- and higher-replica terms.  From rotational invariance it is natural
to look for $\Gamma$ in the form (\ref{action}) with $\vec n_{a} \to
\vec n_a^R=(\sigma_a^R,\vec \pi_a^R)$, $\sigma_{a}\to
\sigma_a^R=\sqrt{1-(\pi_a^R)^2}$, $\pi_{a}\to \pi_a^{R} = Z^{-1/2}
\pi_a$, $T_{0}\to T_R =
 T_0/Z_T$, $M_{0} \to M_R = M_{0} \sqrt{Z}/Z_T$,  $m=\sqrt{M_R}$ the
renormalized mass of the $\vec \pi_a$ modes, and $\hat{R}_{0} (\vec
n_{a} \vec n_{b}) \to m^{\epsilon} \hat{R} (\vec n^{R}_{a} \vec
n^{R}_{b}) $. Higher vertices generated under RG are irrelevant by
power-counting, hence discarded.
Renormalization of $T$ contributes
to the flow of $\hat{R}$, and one sets $T=0$ at the end.

One computes $Z$, $Z_T$ and $\hat R$ perturbatively in $\hat R_0$ and
extracts $\beta$ and $\gamma$ functions $\beta[\hat{ R}](z) = - m \partial_m
\hat{ R} (z)$, $\gamma= - m \partial_m \ln Z$ and $\gamma_{\mathrm{T}}= - m
\partial_m \ln Z_T$, derivatives taken at fixed
$\hat{R}_0,T_{0},M_0$. Although calculation of the $Z$-factors is simplified
due to DR, anomalous contributions appear from the non-analyticity of
$\hat{ R}(z)$. To compute $\hat R(z)$, one chooses a pair of uniform
background fields $(n_a^0,n_b^0)$ for each $(a,b)$. We use a basis for
the fluctuating fields (to be integrated over) such that ${\vec n}_a
=(\sigma_a,\eta_a,\vec \rho_a)$, ${\vec n}_b =(\sigma_b,\eta_b,\vec
\rho_b)$, where $\eta$ lies in the plane common to $(\vec n_a^0,\vec
n_b^0)$, and $\vec \rho_a$ along the perpendicular $N-2$ directions;
both have diagonal propagators. Denoting $\vec n_a^0 \vec n_b^0 = \cos
\phi_{ab}$, one has $\vec n_a \vec n_b = \cos \phi_{ab} \left(
\sigma_a \sigma_b
+ \eta_a \eta_b\right) + \sin\phi_{ab}
\left(\sigma_a \eta_b- \sigma_b \eta_a\right) + \vec \rho_a \vec
\rho_b$.  One gets factors of $(N-2)$ from the contraction of $\vec
\rho$.  Our calculation to 2 loops results in the flow-equation for
the function $R (\phi ) = \hat R(z=\cos \phi)$, and $\epsilon = 4 -d$:
\vspace*{-.4cm}
\begin{widetext} \vspace*{-.5cm}
\begin{eqnarray}
\partial_{\ell} R (\phi ) &=& \epsilon R (\phi )+ \half R''
(\phi)^2-R''(0)R''(\phi) + (N{-}2)\left[\frac 1 2
\frac{R'(\phi)^2}{\sin^2 \phi }-
 \cot \phi R'(\phi)R''(0)\right] \nonumber \\
&& +  \half (R''(\phi)-R''(0) ) R'''(\phi)^2 + 
 (N{-}2) \bigg[ \frac{\cot \phi}{\sin^4 \phi} R'(\phi)^3 
- \frac{5+ \cos 2 \phi}{4 \sin^4 \phi} R'(\phi)^2 R''(\phi)
+ \frac{1}{2 \sin^2 \phi} R''(\phi)^3 \nonumber \\
&& \qquad \qquad \qquad - 
\frac{1}{4 \sin^4 \phi} R''(0) 
\Big( 2 (2 + \cos 2 \phi) R'(\phi)^2 - 6 \sin 2 \phi 
R'(\phi) R''(\phi) + 
(5+ \cos 2 \phi) \sin^2 \phi R''(\phi)^2 \Big) \bigg] \nonumber \\
&& - \frac{N{+}2} 8  R'''(0^+)^2 R'' (\phi )
- \frac{N{-}2}{4} \cot \phi R'''(0^+)^2 R' (\phi )
 - 2 (N{-}2) \Big[R'' (0) - R'' (0)^{2} + \gamma_{a} R''' (0^+)^{2} \Big]
   R (\phi ) \qquad  \label{beta}
\end{eqnarray}
\end{widetext}
\vspace*{-0.9cm} with $\partial_l := - m \partial_m$, and the last
factor proportional to $R (\phi )$ is $- 2 \gamma_{\mathrm{T}}$ and
takes into account the renormalization of temperature. Thanks to the
anomalous terms, arising from a non-analytic $R(\phi)$, this
$\beta$-function preserves a (at most) linear cusp (i.e.\ finite $R'''
(0^+)$), and reproduces for $N=2$ the previous 2-loop results for the
periodic RM \cite{ChauveLeDoussalWiese2000a}. For $N>2$, anomalous
contributions are determined following
\cite{LeDoussalWiese2005}. $\gamma$ is found as
\begin{eqnarray}\label{gamma}
\gamma = (N-1) R''(0) + {\textstyle \frac{3 N-2}{8}} R'''(0^+)^2 \ ,
\end{eqnarray}
either via a calculation of $\langle \sigma_a \rangle$ \footnote{For
$N=2$, one can also use $\eta=\sin \phi$ and a RM calculation with
$\langle \sigma_a \rangle \approx \rme^{- \frac{1}{2} \langle \phi_a^2
\rangle}$, since the field correlator is gaussian up to
$O(\epsilon^4)$ \cite{LeDoussalWieseChauve2003}.}, or of the mass
corrections, a result consistent with the $\beta$ function
(\ref{beta}) \footnote{Reexpressing (\ref{beta}) in $\hat R$,
(\ref{gamma}) is the rescaling term $\gamma z \hat R'(z)$.}. The
determination of $\gamma_{\mathrm{T}}$ is more delicate
\footnote{1-loop corrections to correlations at non-zero momentum are
anisotropic $\sim \mu_{ij}(v)$ (see (13) in \cite{LeDoussalWiese2005})
in presence of a background $\hat{n}_{a}^{0}$ here, $v_{a}$
there. This yields formally $\gamma_{a}^{\eta}=\frac{1}{4}$,
$\gamma_{a}^{\rho}= (3N-4)/ (8 (N-2))$.}, and we have allowed for an
amalous contribution $\gamma_{a}$, whose effect is minor, and
discussed below. The correlation exponents (standard definition
\cite{Feldman2002}) are obtained as $\bar \eta=\epsilon-\gamma$,
$\eta=\gamma_{\mathrm{T}}-\gamma$ at the FP.  (\ref{beta}) has the
form:
\begin{eqnarray}\label{lf9}
\partial_{\ell}R &=&\epsilon R   + {\mathrm B}(R,R) + {\mathrm C}(R,R,R) + 
  O(R^4) 
\end{eqnarray}\begin{figure}[b]
\Fig{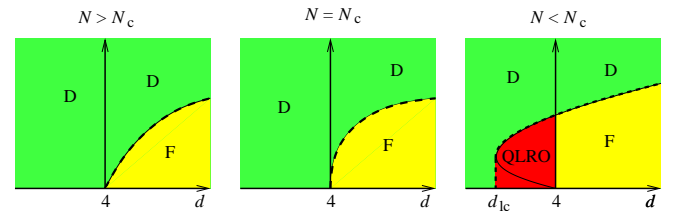}
\caption{(Color online) Phase diagram. D $=$ disordered, F $=$ ferromagnetic,
QLRO $=$ quasi long-range order.}
\label{f:phases}
\end{figure}We now discuss its solution, first in the RF case, and
setting $\gamma_{a}=0$.  The 1-loop flow-equation (setting $C=0$)
admits, in dimensions larger than 4, a fixed point
$R^*_{\mathrm{F/D}}$ with a single repulsive direction, argued by
Feldman to describe the F/D zero temperature transition. This is true
only for $N>N_{c}$.  For $N<N_{c}$ this fixed point {\it disappears}
and instead an {\it attractive fixed point} $R^*_{\mathrm{QLRO}}$
appears which describes the Bragg glass for $N=2$. We have determined
$N_{c}= 2.8347408$ and the solution $R_{c}(u)$ which satisfies
$\mathrm{B}(R_c,R_c)|_{N=N_{c}}=0$. It is formally the solution at
$\epsilon=0$. Since the FRG flow vanishes to one loop along
the direction of $R_c$, examination of the 2-loop terms is needed to
understand what happens at $N=N_c$. In particular the F/D transition
should still exist for $N<N_c$, though it cannot be found at one
loop. It is not even clear a priori whether it remains perturbative.

The scenario found is perturbative, accessible within a double
expansion in $\sqrt{|\epsilon|}$ and $N-N_c$.  To this aim, we write
the leading terms in $N-N_{c}$ and $\epsilon$ of (\ref{lf9}), namely
\begin{eqnarray}\label{lf1}
&&\!\!\!\partial_{\ell}R = \epsilon R + {\mathrm B}_{c}(R,R) + {\mathrm
C}_{c}(R,R,R) +   \left( N {-} N_c \right)  {\mathrm B}_N(R,R)\nonumber \\
&&\!\!\!{\mathrm{B}}_{c} (\dots ) = \mathrm{B} (\dots )|_{N=N_{c}}\ , \qquad
\!\!\!{\mathrm{C}}_{c} (\dots ) = \mathrm{C} (\dots )|_{N=N_{c}}
\end{eqnarray}
One looks for a fixed-point solution of the form $ R (u) = g R_{c} (u)
+ g^{2} \delta R (u)$, with $g>0$, $R_c''(0)=-1$, and its flow. This
analysis is done numerically and leads to the flow shown schematically
on Fig.\ \ref{f:phases}. The RG-flow projected onto the
direction of $g$ is equivalent to
\begin{eqnarray}\label{lf2}
\partial_{l} g &=& \epsilon g + 1.092 (N-N_{c}) g^{2} + 2.352 g^{3}\ .
\end{eqnarray}
As a solution of the functional flow near $N_{c}$, its simplicity is
surprising.  Setting $g = (N_c -N) f$, there are three FP:
\begin{equation}\label{s1}
\frac{\epsilon}{(N-N_{c})^{2}} - 1.092 f + 2.352 f^{2} =0\ , \quad
\mbox{or}\quad  f=0\ .\ \  \label{eqf} 
\end{equation}%
\begin{figure}\setlength{\unitlength}{1.4mm}
\fboxsep0mm 
\mbox{\begin{picture} (60,37)
\put(0,0){\fig{60\unitlength}{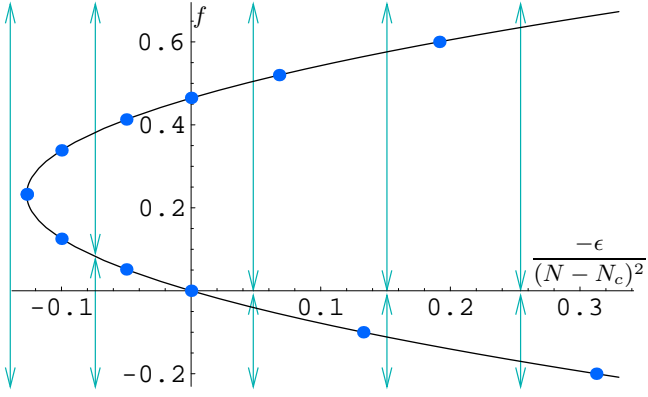}}
\put(50,12){$\displaystyle  \frac{-\epsilon}{(N-N_{c})^{2}}$}
\put(18,35){$f$} 
\end{picture}}
 \caption{(Color online) Parametric plot for solutions of (\ref{lf1})
for $N < N_c$ (dots) for RF, equivalent to (\ref{lf2}) (solid line,
parabola) and flow (arrows). $f$ parameterizes disorder and only $f
\geq 0$ is physical. Cmp. with fig.\ \ref{f:phases}, right.}
\label{f:magnificientRF2}
\end{figure}For $N > N_{c}$ the physical branch is $f<0$. As seen in
Fig.~\ref{f:phases}, for $d > 4$ there is a ferro phase (i.e.\
$f=0$ is attractive) and an unstable FP describing the F/D transition,
given by the negative branch of (\ref{eqf}). At $N=N_c$ one sees from
(6) that the F/D fixed point is still perturbative, but in a
$\sqrt{\epsilon}$ expansion for $g$ (and for the critical exponents).
For $N<N_{c}$ the physical side is $f>0$ and there are two branches on
Fig.~\ref{f:magnificientRF2} corresponding to {\em two} non-trivial
fixed points. One is the infrared attractive FP for weak disorder
which describes the Quasi-Ordered ferromagnetic phase; the second one
is unstable and describes the transition to the disordered phase with
a flow to strong coupling.  These two fixed points exist only for
$\epsilon < \epsilon_{c}$ and annihilate at $\epsilon _{c}$. The
lower critical dimension of the RF-model for $N < N_c$ is lowered from
$d=4$ to
\begin{equation}\label{lf4}
d_{\mathrm{lc}}^{\mathrm{RF}} = 4-\epsilon_{c}\approx 4 - 0.1268
(N-N_{c})^{2}+ O ( (N-N_{c})^{3})\ .
\end{equation}
Note that the mechanism is different from the more conventional
criterion $d-4+\eta(d)=0$ at $d=d_{\mathrm{lc}}$.

The same analysis for the random anisotropy class yields
$N_{c}=9.44121$. The equivalent of (\ref{lf2}) becomes $\partial_{l} g
= \epsilon g + 0.549 (N-N_{c}) g^{2} + 47.6 g^{3}$, leading to
$d_{\mathrm{lc}}^{\mathrm{RA}} \approx 4-0.00158 (N-N_{c})^{2}
$. Although it yields $d_{\mathrm{lc}}(N=3) \approx 3.93$ and no QLRO
phase in $d=3$, naive extrapolation should be taken with caution given
the high value of $N_c$.
Numerical values for $d_{\mathrm{lc}}$ are changed for a
 $\gamma_{a}\neq 0$, but  the scenario is robust for
$\gamma_{a}< \gamma_{c}$ \footnote{The coefficient
 in (\ref{lf2}) and (\ref{eqf}) becomes
 $2.35(1-\gamma_{a}/\gamma_{c})$ with $\gamma_{c}= 2.04$,
and similarly
 for RA $47.6 (1-\gamma_{a}/\gamma_{c})$, $\gamma_{c}=1.23$,
with a corresponding shift $4-d_{\mathrm{lc}}(\gamma_a) =
(4-d_{\mathrm{lc}}(\gamma_a=0))/(1-\gamma_{a}/\gamma_{c})$.  For $\gamma_{a}
>\gamma_{c}$ the scenario reverses (fig.\
 2 is flipped w.r.t.\ the $f$-axis). 
The FP, for $\epsilon <0$ and $N<N_{c}$ are  at $\epsilon
>0$ and $N>N_{c}$. The bifurcation  occurs entirely within the
Ferro phase, and the QLRO branch survives above $d=4$. This scenario
would imply a F/D fixed point inaccessible to  FRG,
contradicting  \cite{Feldman2002}. It is unlikely,
since [32] suggests $\frac{1}{4}\le \gamma_{a}\le \frac{1}{2}$.}.

\begin{figure}[t]\setlength{\unitlength}{1.2mm}
\fboxsep0mm 
\mbox{\begin{picture} (64,37)
\put(4,0){\fig{57\unitlength}{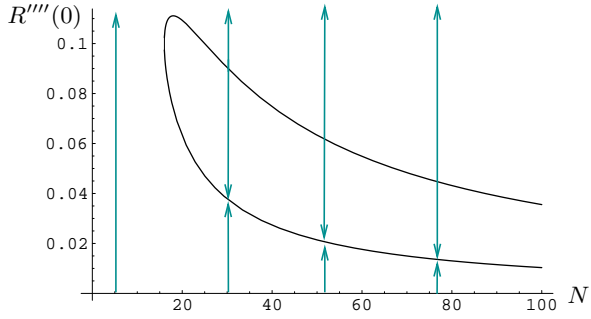}}
\put(62,2){$N$}
\put(0,33){$R'''' (0)$} 
\end{picture}}
 \caption{Flow for $R'''' (0)/|\epsilon|$ for $d=4-\epsilon > 4$ as a function of $N$.
The two branches behave as $1/N$ and $4/N$  at large $N$.
}
\label{f:R4flow}
\end{figure}

We now discuss the FRG flow-equations for $N$ large. From a truncated
exact RG Tarjus and Tissier (TT) \cite{TarjusTissier2004} found: that
the linear cusp of the F/D fixed point for $d>4$ vanishes for $N >
N^*(d)$, i.e.\ $R'''(0^+)=0$; and that the non-analyticity becomes
weaker as $N$ increases (as $|\phi|^n$ with $n \sim N$).  Analytical
study of the derivatives of (\ref{beta}) confirms the existence of
this peculiar FP to two loop and predicts $N^*(d,2p)$ beyond which the
set of $\{ R^{(2k)}(0) \}$ for $k \leq p$ admits a stable FP, with
$R^{( 2k-1)}(0^+)=0$ for $k \leq p$ and $R^{( 2k-1)}(0^+) \neq 0$ for
$k > p$. We find:
\begin{equation}
N^*(d) = N^*(d,4) = 18 + 49\epsilon/5  + \dots 
\end{equation}
which yields a slope roughly twice the one of Fig.~1 of
\cite{TarjusTissier2004}. This remarkable FP raises some
puzzles. Although weaker than a cusp its non-analyticity should imply
some (weaker) metastability in the system. It is thus unclear whether
DR is fully restored: to prove it one
should rule out feedback from anomalous higher-loop terms in exponents
or the $\beta$-function. Finally, one also wonders about its basin of
attraction. As shown in Fig.~\ref{f:R4flow}, the FRG flow for
$R''''(0)$ is still to large values if its bare value is large enough,
indicating some tendency to glassy behaviour.

To explore these effects we now study the F/D phase
transition at large $N$ and $d>4$. We
obtain, both at large $N$ and fixed $d$ (extending
Ref.~\cite{LeDoussalWiese2001}), and to one loop, the flow equation
for the rescaled $\tilde R(z=\cos \phi) = N R (\phi)/|\epsilon|$:
\begin{equation}\label{lf148}
\partial_l \tilde R = - \tilde R + 2 \tilde R_{1}' \tilde R - \tilde
R_{1}' \tilde R' z + {\textstyle \half} \tilde R'^{2} = 0 \ .
\end{equation}
We denote $y(z)=\tilde R'(z)$, $y_0=\tilde R'(1)=-N R''(0)/|\epsilon|$
and $r_4=N R''''(0)/|\epsilon|$. There are two analytic FPs $\tilde
R(z)=z-1/2$ and $\tilde R(z)=z^2/2$, corresponding both to $y_0=1$ and
to $r_4=1$ and $r_4=4$ respectively.  This agrees with the flow of thed
erivatives for analytic $R(\phi)$: $\partial_l y_0 = y_0 (y_0 -1)$,\
and at $y_{0}=1$: $\partial_l r_4 = \frac{1}{3} (r_4-1)(r_4-4) $.  The
first FP is the large-$N$ limit of the TT fixed point, the second is
repulsive and divides the region where $r_4\to \infty $ (non-analytic
$R(\phi)$) in a finite RG time $l_c$ (Larkin scale). For $y_0>1$, we
find a family of NA fixed points with a linear cusp, parameterized by
an integer $n\ge 2$, s.t.\ $y_0=n/ (n-1)$, $z= y - (y_{0}-1)
(y/y_{0})^{n}$.  The solutions with $n$ (i.e.\ $z(y)$) odd correspond
to random anisotropy ($R'(\phi)=R'(\phi+\pi)$). The $n=2$ RF fixed
point is $R (\phi)=2 \cos (\phi)+ \frac{8 \sqrt{2}}{3} \sin^3
(\phi/2) - \frac{4}{3}$. To elucidate their role, we obtained the
exact solution for the flow both below $l_c$, i.e.\ $z = \frac{y}{y_0}
+ (y_0-1) \Phi(\frac{y}{y_0})$ ($\Phi(x)$ parameterizes the bare
disorder, $\Phi(1)=0$), and above $l_c$, with an anomalous flow for
$y_{0}$. Matching at $l_c$ yields the critical manifold for RF
disorder, defined from the conditions that $\Phi'(w)=\Phi(w)/w=1$ has
a root $0\le w\le 1$.  It is different
from the naive DR condition $y_0=1$, valid for small $r_4$. The $n=2$
FP corresponds to bare disorder such that the root $w=0$. Hence it is
multicritical \footnote{We thank G.\ Tarjus and M.\ Tissier for pointing out
this important fact.}. Generic initial conditions within the critical
F/P manifold flow back to the TT FP i.e.\ the linear cusp decreases to
zero \footnote{This is true only on one side of the multicritical
FP. The other side, if accessible from physically realizable bare
disorder, would correspond to a strong disorder regime.}. This however
occurs only at an infinite scale, hence we expect a long crossover
within a glassy region, characterized by a cusp, and metastability on
finite scales \footnote{The physics associated to a similar reentrant
crossover for RM for $d>4$ is discussed in Appendix H of: L. Balents
and P.~{Le Doussal}, Annals of Physics {\bf 315} (2005) 213.}. The
large-$N$ limit here is subtle. Taking $N\to \infty$ at fixed volume
on a bare model with $\hat R_0(z) = z$ yields only the analytic FP,
equivalent to a replica-symmetric saddle point. Higher monomials $z^p$
are generated in perturbation theory, at higher order in $1/N$. Thus,
for $N$ large but fixed and infinite size, one must first coarse grain
to generate a non-trivial function $\hat R_0(z)$, before taking the
limit of $N\to \infty$.

In conclusion we obtained the 2-loop FRG functions for the random
field and anisotropy $\sigma$-models. We found a new fixed point and a
scenario for the decrease of the lower critical dimension. This rules
out the scenario left open at one loop that the bifurcation close to
$d=4$  simply occurs within the (quasi-) ordered phase.

\end{document}